\documentstyle[11pt]{article}
\newcommand{\nc}{\newcommand}
\newcommand{\dis}{\displaystyle}

\newcommand{\bi}{\bibitem}

\nc{\erm}{{\rm e}}

\begin{document}

\title{{\small{\em Published in} J.\,de Phys.\,(France)\,{\bf I4}\,(1994)\,1783}\\
COMPLEX--BARRIER TUNNELLING TIMES\footnote{Work partially supported by
I.N.F.N.}}

\author{FABIO \ RACITI and GIOVANNI \ SALESI}

\date{}
\maketitle
\begin{center}
{{\em Dipartimento di Fisica, Universit\`a Statale di Catania, Catania,
Italy};\\
and\\
{\em Istituto Nazionale di Fisica Nucleare--Sezione di Catania, Catania,
Italy$^{\star}$}}\\
\end{center}
\footnotetext{$\!\!\!\!^{\star}\,$e-mail: {\em fraciti@dmi.unict.it}; \
{\em salesi@ct.infn.it}}

\vspace{0.5 cm}

\begin{abstract}
In this paper we calculate the analytic expression of the {\em phase
time} for the  scattering of an electron off a complex square barrier. As is
well known the (negative) imaginary part of the potential takes into account,
phenomenologically, the absorption. \ We investigate the Hartman-Fletcher
effect, and  find that it is suppressed by the presence of a (not negligible)
imaginary potential. In fact, when a sufficiently large absorption is present,
the asymptotical transmission speed is finite. Actually, the tunnelling time
does increase linearly with the barrier width. A recent optical experiment
seems to be in agreement with our theoretical expectation.

\end{abstract}

\section{Introduction}

In recent times the longstanding question of the tunnelling times
has acquired new urgency because of the recent experimental results claiming
for superluminal tunnelling speeds. The problem of defining tunnelling
times has a long history, since it arose in the forties and
fifties$^{[1],[2],[3]}$ simultaneously with the fundamental
problem of introducing {\em time} as a  quantum mechanical observable and,
in particular,  of a definition (in Quantum Mechanics) of the collision
durations. Furthermore, a corpuscular picture of tunnelling is very hard to be
realized because of the lack of a direct classical limit
for the sub-barrier particle paths and  velocities.
Nevertheless, various typical definitions for the
time spent by a particle in the classically forbidden regions have been
proposed$^{[4],[5]}$; we underline that the  differences among them are
not only merely formal but, on the contrary, they are a consequence
of different views, physical interpretations and experimental
expectations. \ Let us quote, incidentally, the most important definitions
of the tunnelling times:

a) the dwell time$^{[4],[5],[6]}$: i.e., the time spent inside the barrier, averaged
over all the incoming particles,with no distinction between  transmission
and reflection channels;

b) the local ``Larmor times"$^{[6]}$: i.e., the traversal time as measured
by the spin precession of the tunnelling particle in a uniform infinitesimal
magnetic field;

c) the ``complex time approach"$^{[5],[7]}$: a quantum extension ---via
path integral averages over the classical paths--- of the classical complex
time spent by the particle in the scattering process;

d) the ``Buttiker-Landauer" times$^{[5],[6],[8]}$: namely interaction
times of the particle with a time modulated barrier;

e) the so called ``spatial approaches"$^{[5],[9]}$, based on the probabilistic
quantum standard interpretation of the flux densities $J[x,t]$ involved
during transmission and reflection;

f) the ordinary ``phase times",$^{[1],[2],[4],[5],[10]}$ or group delays:
i.e.,the times taken by quasimonochromatic  wave packets to appear on the
other side of the barrier, as given by the stationary phase approximation.

\

\ Before going on, let us stress the important fact that many of those
theoretical pictures do imply the so-called Hartman-Fletcher effect$^{[10]}$,
that is to say the surprising occurrence, for sufficiently opaque barriers,
of {\em tunnelling delays independent of the barrier width}.\ Namely, those
delays  imply traversal mean group velocities larger than the light speed in
vacuum. \ Recent optical experiment seem to confirm the exitence of superluminal
tunnelling speeds$^{[11]}$. They can be grouped in 3 main classes:

1) evanescent  wave propagation in a low dielectric costant
region, separating two regions of higher dielectric costant or, similarly,
in optical devices allowing for frustrated total internal reflection;

2) propagation of a gaussian light pulse through an anomalous dispersion
medium;

3) evanescent microwaves in a wave guide below cut-off.

\

\ We are particularly interested in evanescent microwaves since the
recent  optical experiments by Nimtz et al.$^{[12]}$ agree with our quantum
mechanical calculation. The analogy between quantum tunnelling and
propagation of microwaves in wave guides is based on the fact that the
group velocity is obtained from the derivative of the trasmission amplitude
in both cases. Moreover the Schroedinger equation is formally identical
to the Helmholtz equation for the propagation of a scalar field, electric or
magnetic component of the wave:
$$
d^{2}\psi/dx^{2} +k^{2}\psi = 0
$$
where $k$ is the wave number. Thus is possible to simulate quantum tunnelling
studying the propagation of e.m. waves in wave guide, where the region below
the cut off is the equivalent of the quantum barrier$^{[11]}$. The connection between
the Schroedinger equation and the Helmholtz equation is also valid for
the Dirac equation, which is a relativistic equation. Using the Dirac
equation instead of the Schroedinger equation we obtain again the Hartman
effect, so that this effect is not  a ``wrong" quantum mechanical effect
due to the fact  that one is not using a relativistic equation (this has
been showed by C.R. Leavens).

Our proposal in this paper is  studying the tunnelling time and the
Hartman-Fletcher effect in the presence of absorption. To that purpouse we
introduce a complex square potential (with a negative immaginary part); such
non-real potentials are customary in scattering theory and in nuclear
physics (where they are named {\em optical potentials}).
We shall show that the non reality of the effective hamiltonian (related to
the existence of other interaction channels) does in general destroy the
Hartman-Fletcher effect. Superluminal speeds can be achieved only if
the imaginary part of the hamiltonian is chosen sufficiently small. Let
us notice that these theoretical predictions are in agreement with the
experimental results obtained by Nimtz et al. at Cologne$^{[12]}$  employing
evanescent microwaves in  absorptive wave guides below cut-off.\ In this
context, the most suitable and natural theoretical approach for the evaluation
of the {\em global} tunnelling times and of the mean tunnelling speeds is
perhaps the most ``direct" one, the phase time approach already mentioned in
f): thus, in the following we are going to adopt such a theoretical approach.

\

\section{Analytic calculation of the phase time}

\ Following the ordinary procedures employed in refs.$[10]$, the group
delay for quasimonochromatic packets in the stationary phase approximation,
is given by:
$$
\delta\tau (E) =\hbar{\partial\over\partial E} (\arg A_{\rm T})
\eqno(1{\rm a})
$$
where $E$ is the incoming particle energy, $A_{\rm T}$ is the (complex)
transmission amplitude and $\hbar$ is the reduced Planck constant.
The global tunnelling time is given, as usual, by the sum of the semiclassical
traversal and delay time:
$$
\tau = \tau_{\rm c} + \delta\tau (E)\eqno(1{\rm b})
$$
In order to find out the analytic expression for $A_{\rm T}$ we have to solve
the (stationary) Schroedinger equation with a potential different from zero
only in the interval $(0,a)$; namely:
$$
V(x) = V_{0}-iV_{1}, \, \, \; \; \;{ x\epsilon(0,a)} \; .
$$
Let us observe that:
$$
\psi_{\rm I} \equiv \psi_{\rm in} + \psi_{\rm R}  =  \erm^{ikx} +
B_{\rm I} \erm^{-ikx}
$$
and
$$
\psi_{\rm III} \equiv \psi_{\rm T} = A_{\rm III} \erm^{ikx}
$$
as in the real potential case (with $k^{2} = 2mE/{\hbar^{2}}$),
while $\psi_{\rm II}$ is obtained by solving the Schroedinger equation
in the barrier region:
$$
\frac{{\rm d^{2}}\psi}{{\rm d}x^{2}} + {2m\over  \hbar^{2}} (E-V_{0} +
iV_{1}) \psi = 0 \eqno(3)
$$
Thus, we get for $\psi_{\rm II}$ the following expression:
$$
\psi_{\rm II} = A_{\rm II} \erm^{ik_{\rm II} x} + B_{\rm II}
\erm^{-ik_{\rm II} x} \eqno(4)
$$
where
$$
k_{\rm II} \equiv \sqrt{2m(E-V_{0} +iV_{1})}/\hbar\eqno(5{\rm a})
$$
Let us notice that  eq.(5a) implies that $k_{\rm II}$ is a complex quantity:
$$
k_{\rm II} \equiv \xi + i \mu\eqno(5{\rm b})
$$
where $\xi, \mu$ are real numbers.

In the $E < V_{0}$ case we get:
$$
\xi = (\sqrt{m}/\hbar)\sqrt{\sqrt{(E-V_{0})^{2} +V_{1}^{2}} -(V_{0}-E)}
\eqno(6{\rm a})
$$
and:
$$
\mu = (\sqrt{m}/\hbar)
\sqrt{\sqrt{(E-V_{0})^{2} +V_{1}^{2}} +(V_{0}-E)}\eqno(6{\rm b})
$$
Imposing the continuity boundary conditions for $\psi$ and its derivative,
i.e.:
$$
\psi_{\rm I} (0) = \psi_{\rm II} (0) \, ; \ \ \psi_{\rm II} (a) =
\psi_{\rm III}(a) \, ; \ \ \psi^{\prime}_{\rm I} (0) =
\psi^{\prime}_{\rm II} (0) \, ; \ \ \psi^{\prime}_{\rm II}(a) =
\psi^{\prime}_{\rm III} (a) \; ,
$$
we get, after some algebra,
the expression for $A_{\rm T} \equiv A_{\rm III}$ :
$$
A_{\rm T} = {4kk_{\rm II} \erm^{ik_{\rm II}a}\erm^{-ika} \over
{k^{2} +{k_{\rm II}^{2}}(1-\erm^{2ik_{\rm II}a}) +
2kk_{\rm II}(1 + \erm^{2ik_{\rm II}a})}}\eqno(7{\rm a})
$$
Since we are interested in determining $\phi\equiv \arg A_{\rm T}$
let us express $A_{\rm T}^{-1}$ in agebraic form, e.g.:
$$
A_{\rm T}^{-1} = \frac{Ax+By+C\omega}{2kk_{\rm II}k_{\rm II}^{*}} +
i\frac{Cr-Ax+By}{2kk_{\rm II}k_{\rm II}^{*}}\eqno(7{\rm b})
$$
where:
$$A \equiv \xi(k^{2}+k_{\rm II}k_{\rm II}^{*})\eqno(8{\rm a})$$
$$B\equiv {\mu}(k_{\rm II}k_{\rm II}^{*} -k^{2})\eqno(8{\rm b})$$
$$C \equiv 2k k_{\rm II}k^{*}_{\rm II}\eqno(8{\rm c})$$
$$x \equiv {\sin}{{\xi} a\; {\cosh}{\mu} a}\eqno(8{\rm d})$$
$$y \equiv {\cos}{\xi}{a}\;{\sinh}{\mu}a\eqno(8{\rm e})$$
$$\omega\equiv {\cos}{{\xi}{a}}\; {\cosh}{{\mu}a}\eqno(8{\rm f})$$
$$r\equiv -{\sin}{{\xi}a}\; {\sinh}{{\mu}{a}}\eqno(8{\rm g})$$
Thus we may write:
$$
\phi \equiv \arg A_{\rm T} = \arctan(\frac{-Cr+Ax-By}{C\omega+Ay +By})-
ka.\eqno(9{\rm a})
$$
After some  manipulations, quantity $\phi$ can be written
as follows:
$$
\phi =\arctan(\frac{\tanh{\mu}a(C-B\cot{\xi}a)+A}{\cot{\xi}a(C+
A{\tanh{\mu}a}) +B}) -ka.\eqno(9{\rm b})
$$
According to eq. (1b) the tunnelling time is:
$$
\tau(E) = {a \over v}
+\hbar{\partial \over \partial{E}} (\arg{A_{\rm T}}) =
\hbar{\partial \over {\partial
E}}(\arg{A_{\rm T}+ ka}) \eqno(10)
$$
where $v=\hbar k/m$. By inserting eq. (9b) into eq. (10) we get, after some
elaborations (see the Appendix),
$$
\tau = {n \over d} \eqno(11)
$$
where
$$
n \equiv {\sin}{2\xi}a[(-am\mu/{{\hbar^{2}}{\rho^{2}}})(C^{2}-B^{2}-A^{2})+
(A^{\prime}C-AC^{\prime})] \; + \; \cos2{\xi}a [{2BCam\mu/({\hbar^{2}}
{\rho}^{2})}] \; +$$
$$+ \, {\sinh{2\mu}a[(am\xi/\hbar^{2} \rho^{2})(A^2+B^2+C^2)+
(BC^{\prime}-B^{\prime}C)] \; + \;
(2am{\xi}AC/{\hbar^{2} \rho^{2}}) \cosh{2\mu}a} \; +$$
$$+ \; 2(A^{\prime}B-AB^{\prime})({\sin^{2}{\xi}a}+{\sinh^{2}{\mu}a})
\eqno(11{\rm a})
$$
and
$$
d \equiv 2(A^2+B^2)({\sin^{2}{\xi}a}+{\sinh^{2}{\mu}a})+
2AC\sinh2{\mu}a +2BC\sin2{\xi}a+
2C^2({\cos^{2}{\xi}a}+{\sinh^{2}{\mu}a}). \eqno(11{\rm b})
$$
Since we want to check the occurrence of the Hartman-Fletcher
effect, we are interested in studying the opaque barrier limit.
That is, we are interested in the asymptotic condition:
$$
ak_{\rm II}k_{\rm II}^{*} >> 1\eqno(12)
$$
which yields:
$$
\tau^{\rm asy} = \frac{m {\xi} a}{\hbar (\xi^{2} + \mu^{2})}\eqno(13)
$$
that is, $\tau^{\rm asy}$ is directly proportional to the barrier width. In other
words the mean tunnel speed $v_{l}$ is asymptotically :
$$
v_{l} \equiv{ a\over \tau^{\rm asy}} = \frac{\hbar(\xi^{2}+\mu^{2})}{
m \xi}.\eqno(14)
$$
Thus, we do not obtain the Hartman-Fletcher effect and the saturation
of the transmission times anylonger, but a limiting speed as it is shown
in eq. (14). Nevertheless, as we can deduce from (14), for sufficiently small
values of ${\xi}$,i.e., in the case of low absorption, we get:
$$
v_{l} \,\longrightarrow \,{\infty}
$$
so that superluminal tunnelling velocities are not {\em a priori} forbidden.

\

\section{Conclusions}

\ Most of the theoretical models proposed for the tunnelling time
give rise to the Hartman effect, i.e. to an anomalously short
tunnelling time. In the very last years the phase time approach,
as well as the Hartman effect, has received experimental confirmation
by several groups which exploited the analogy between tunnelling particles
and evanescent electromagnetic waves. Thus, we propose to extend this approach
to the case of a complex potential, in order to introduce an absorption
channel. Our calculations  shows that a strong absorption tends to destroy
the Hartman effect and agree with the experimental results of Nimtz et al.
\ However, although all those experiments do confirm the Hartman effect,
there are controversial opinions about the interpretation of this
effect and the possibility of  superluminal signal transmission.
\ In our opinion, for a better physical interpretation of this
phenomenus  one should also study the {\em distributions}
of tunnelling times and the distortion of the wave packet. Work
is in progress in this direction.

\

\

\ {\bf Acknowledgements}

\noindent Special thanks are due to V.S. Olkhovsky for having suggested this work,
and to E. Recami and V.S. Olkhovsky for stimulating discussions and
their continuous help. \ We also thank S. Sambataro, R. Mignani and
G. Andronico, G. Angilella, P. Falsaperla, A.Lamagna, G.D. Maccarrone,
R. Maltese. \ At last, the kind collaboration of G. Nava, S. Parietti, P. Saurgnani
and T. Venasco is also acknowledged.

\

\

\ {\bf APPENDIX}\\

Our starting point will be  the analytical expression for $\phi$,
already written in (9b). Thus, we can calculate the global tunnelling
time (10) and obtain that in the general case:
$$
{\partial\over\partial E}({\phi+ka}) ={n \over d}
$$
where
$$
n \equiv (1/\cosh^{2}\xi{a}) \{ \sin{\xi}a\cos{\mu}a
[a\mu^{\prime}(C^{2}-B^{2}-
A^{2})+(A^{\prime}C-AC^{\prime})] \; + \; a\mu^{\prime}BC(\sin^{2}{\xi}a-
\cos^{2}{\xi}a) \} \; +$$
$$+ \; \tanh{\mu}a[a{\xi}^{\prime}(A^{2}+B^{2}+C^{2})+
(BC^{\prime}-B^{\prime}C)] \; +$$
$$+ \; a{\xi}^{\prime}AC(1+\tanh^{2}{\mu}a)+
(A^{\prime}B-AB^{\prime})(\sin^{2}{\xi}a+\tanh^{2}{\mu}a\cos^{2}{\mu}a)
$$
and
$$
d \equiv (A^{2}+B^{2})(\sin^{2}{\xi}a+\tanh^{2}{\mu}a\cos^{2}{\xi}a) \; +
\; 2AC\tanh{\mu}a \; + \; C^{2}(\cos^{2}{\xi}a+\sin^{2}{\xi}a\tanh^{2}{\mu}a)
\; +$$
$$+ \; {\dis{\frac{BC\sin2{\xi}a}{\cosh^{2}{\mu}a}}} \; ,
$$
symbol {$\prime$} meaning derivation with respect to the energy.

At this point, we can check our complicated formula in the special case
$V_{1}=0$. \ Let us observe that the assumption $V_{1} =0$ implies:
$$
\xi=0, \ \ \xi^{\prime}=0, \ \ {\mu}^{\prime}=-{m/\hbar^{2}}(1/\mu) \; ;
\ \  k^{\prime}={m/\hbar^{2}}(1/k) \ ;
$$
and also:
$$
A = 0 \; ; \ \ A^{\prime} = 0 \; ; \ \
B={\mu}^{3} -{\mu}k^{2} \; ; \ \ B^{\prime}=3{\mu}^2{\mu}^{\prime}-{\mu}^
{\prime}k^2-
2{\mu}kk^{\prime} \; ;$$
$$C=2k{\mu}^{2} \; \ \ C^{\prime}=2{\mu}^{2}k^{\prime}+4{\mu}{\mu}^{\prime}
k \; .
$$
After some manipulations,  we get the expression for the phase time in the
particular case $\;$ $V_{1}=0$:
$$
{\partial \over {\partial E}} {(\phi+ka)}_{V_{1}=0} = \frac{{(am/\hbar^{2})2k^{2}\mu
(\mu^{2}-k^{2}) +(m/\hbar^2)(2mV_{0}/\hbar^{2})^{2})\sinh{2\mu}a}}
{{k\mu[(2m/\hbar^{2})V_{0}^{2}{\sinh{\mu}a}^{2}+4k^{2}\mu^{2}}} \ .
$$
If we define \ $(2mV_{0}/\hbar^{2})^{2}\equiv {k_{0}^{2}}$ \
and \ $D\equiv [(2m/\hbar^{2})V_{0}^{2}{\sinh{\mu}a}^{2}+4k^{2}\mu^{2}]$, \
we realize that our expression is identical with eq.(12a) of ref.[5].
After this check, we can now go back to the general expression (11)
of the tunnelling time which ---after further algebra--- can be written:
$$
\tau = {n \over d}
$$
where now
$$
n \equiv {\sin}{2\xi}a[(-am\mu/{{\hbar^{2}}{\rho^{2}}})(C^{2}-B^{2}-A^{2})+
(A^{\prime}C-AC^{\prime})]+\cos2{\xi}a [{2BCam\mu/({\hbar^{2}}{\rho}^{2})}]
\; +$$
$$+ \; {\sinh{2\mu}a[(am\xi/\hbar^{2} \rho^{2})(A^2+B^2+C^2)+(BC^{\prime}
-B^{\prime}C)]+
(2am{\xi}AC/{\hbar^{2} \rho^{2}}) \cosh{2\mu}a} \; +$$
$$+ \; 2(A^{\prime}B-AB^{\prime})({\sin^{2}{\xi}a}+{\sinh^{2}{\mu}a})
$$
and
$$
d \equiv 2(A^2+B^2)({\sin^{2}{\xi}a}+{\sinh^{2}{\mu}a})+
2AC\sinh2{\mu}a +2BC\sin2{\xi}a \; +$$
$$+ \; 2C^2({\cos^{2}{\xi}a}+{\sinh^{2}{\mu}a}) \ .
$$

\

\

\end{document}